\documentclass[pra,showpacs,amsmath,amssymb,twocolumn, 10pt]{revtex4}

\usepackage{amsthm}
\usepackage{dcolumn}
\usepackage{bm}
\usepackage{graphicx}

\begin{document}

\newtheorem{corollary}{Corollary}
\newtheorem{definition}{Definition}
\newtheorem{example}{Example}
\newtheorem{lemma}{Lemma}
\newtheorem{proposition}{Proposition}
\newtheorem{theorem}{Theorem}
\newtheorem{fact}{Fact}
\newtheorem{property}{Property}
\newcommand{\bra}[1]{\langle #1|}
\newcommand{\ket}[1]{|#1\rangle}
\newcommand{\braket}[3]{\langle #1|#2|#3\rangle}
\newcommand{\ip}[2]{\langle #1|#2\rangle}
\newcommand{\op}[2]{|#1\rangle \langle #2|}

\newcommand{\tr}{{\rm tr}}
\newcommand {\E } {{\mathcal{E}}}
\newcommand {\F } {{\mathcal{F}}}
\newcommand {\diag } {{\rm diag}}

\title{Optimal Simulation of a Perfect Entangler}
\author{Nengkun Yu}
\email{nengkunyu@gmail.com}
\author{Runyao Duan}
\email{dry@tsinghua.edu.cn}
\author{Mingsheng Ying}
\email{yingmsh@tsinghua.edu.cn}

\affiliation{State Key Laboratory of Intelligent Technology and
Systems, Tsinghua National Laboratory for Information Science and
Technology, Department of Computer Science and
Technology,Tsinghua University, Beijing 100084, China and\\
Center for Quantum Computation and Intelligent Systems, Faculty of
Engineering and Information Technology, University of Technology,
Sydney, NSW 2007, Australia}

\date{\today}

\begin{abstract}
A $2\otimes 2$ unitary operation is called a perfect entangler if it
can generate a maximally entangled state from some unentangled
input. We study the following question: How many runs of a given
two-qubit entangling unitary operation is required to simulate some
perfect entangler with one-qubit unitary operations as free
resources? We completely solve this problem by presenting an
analytical formula for the optimal number of runs of the entangling
operation. Our result reveals an entanglement strength of two-qubit
unitary operations.
\end{abstract}


\maketitle A fundamental problem in quantum computation is to
understand what kind of quantum resources can be used to accomplish
universal quantum computation, i.e., can be used to simulate any
other quantum circuit either approximately or exactly. Due to its
significance, a great deal of research works have been done in the
last two decades (See Chapter 4 of \cite{NC03} for an excellent
review). For instance, it is now clear that any fixed entangling
two-qubit unitary operation(or Hamiltonian) together with all
one-qubit unitary operations is exactly universal
\cite{BB01,BDDG+02}. Notably, the minimal time of simulating a
two-qubit unitary operation with a given two-qubit Hamiltonian
together with local unitary operations has been obtained
\cite{HVC02,CHN03}. However, the minimum number of runs to simulate
a two-qubit unitary operation using a fixed two-qubit unitary
operation and local unitary operations remains unknown.

In practice we need to simulate unitary operations with some
prescribed properties rather than arbitrary ones. In particular, we
are interested in clarifying bipartite unitary operations according
to their ability of generating entanglement \cite{ZZF00,KC01,
BHLS03,YSZG04}. We call a two-qubit unitary operation a perfect
entangler if it can transform an initially unentangled input into a
maximally entangled pure states by a single run \cite{KC01}. This is
different from the stronger notion of universal entangler introduced
in \cite{CDJY+08}. Obviously, the class of perfect entanglers is
very in quantum information processing. The structure of perfect
entanglers has been thoroughly characterized in Refs. \cite{KC01,
ZVSW03}. An interesting question is thus to ask how many runs of a
fixed unitary operation are required in order to simulate some
perfect entangler, providing that local unitary operations are free
resources.

The purpose of this paper is to provide an analytical solution to
the above question. From another viewpoint, we have obtained the
minimum number of runs of a fixed two-qubit unitary operation
required to create a maximally entangled state from an unentangled
product state. Interestingly, the optimal number of runs is
determined by a single quantity which can be easily calculated from
the nonlocal parameters of the given two-qubit unitary operation.
Our finding reveals some new unexpected structure of two-qubit
unitary operations. Furthermore, our proof techniques can be used to
provide some nontrivial lower bounds between two-qubit unitary
operations.

A by-product in our proof is that for two-qubit unitary operation
$U$ the number of runs of $U$ required to transform a product state
into a maximally entangled state is the same as the number of runs
of $U$ required to transform a maximally entangled state into a
product state, where we assume the local unitary operations are free
resources. This can be understood as a generalized version of the
result that for two-qubit unitary operation $U$ the entangling power
is the same as the disentangling power \cite{BS03}. Note that the
entangling power and disentangling power of a unitary operation $U$
are not always equal in higher-dimensional case as it has been shown
that there is $2\otimes 3$ unitary operation $U$ such that in the
presence of ancillas the entangling power of $U$ is not equal to the
disentangling power of $U$ \cite{LSW05}, and the gap may be of
$O(\log d)$ for some very special $d\otimes d$ unitary operation
$U$.

Throughout this paper, consideration will be restricted to two-qubit
systems. Let us begin with some preliminaries that are useful in
presenting our main results.  We will use the magic basis consisting
of the following states \cite{Woo98}:
$|\Psi_1\rangle=(|00\rangle+|11\rangle)/\sqrt{2},
|\Psi_2\rangle=i(|00\rangle-|11\rangle)/\sqrt{2}$,
$|\Psi_3\rangle=(|01\rangle-|10\rangle)/\sqrt{2},
|\Psi_4\rangle=-i(|01\rangle+|10\rangle)/\sqrt{2}.$ The employed
measure of entanglement is the concurrence introduced by Wootters
\cite{Woo98}. Let $\ket{\psi}$ be a $2\otimes 2$ state such that
$\ket{\psi}=\sum_{k=1}^4 \mu_k\ket{\Psi_k}$. Then the concurrence of
$\ket{\psi}$ is given by
\begin{equation}
C(\ket{\psi})=|\sum_{k=1}^4 \mu_k^2|,\end{equation}

Then $0\leq C \leq 1$, especially, $C(\ket{\psi})$ attains one if
and only if $\ket{\psi}$ is maximally entangled, which means that
$\mu_k$ are all real up to some phase factor; $C(\ket{\psi})$
vanishes if and only if $\ket{\psi}$ is a product (unentangled)
state.

An extremely useful tool in studying two-qubit unitary operations is
a canonical decomposition of two-qubit unitary operations. More
precisely, each two-qubit unitary operation $U$ can be expressed
into the following way ~\cite{KC01}:
\begin{equation}
U=(u_A\otimes u_B)U_d(v_A\otimes v_B),
\end{equation}
where $u_A$, $u_B$, $v_A$, $v_B$ are one-qubit unitary operations,
and $U_d=\exp(i(\alpha_x \sigma_x\otimes \sigma_x+\alpha_y
\sigma_y\otimes \sigma_y+\alpha_z \sigma_z\otimes \sigma_z))$, and
$\sigma_X$, $\sigma_y$, $\sigma_z$ are Pauli matrices. In other
words, every $2\otimes 2$ unitary operation $U$ is equivalent to a
special form of $U_d$ up to some local unitary operations. Most of
the nonlocal properties of $U$ are essentially determined by $U_d$.

Note that $\sigma_x\otimes \sigma_x$, $\sigma_y\otimes \sigma_y$,
and $\sigma_z\otimes \sigma_z$ are pairwise commutative, and thus
have a set of common eigenvectors $\{\ket{\Psi_k}:1\leq k\leq 4\}$.
We can diagonalize $U_d$ as follows:
\begin{equation}\label{Ud2}
U_d=\sum_{k=1}^4 e^{i\lambda_k}\op{\Psi_k}{\Psi_k},
\end{equation}
For a unitary operation $U$, we denote $\Omega(U)=\Theta(U_d^2)$,
where $\Theta(U)$ denotes the length of the smallest arc containing
all the eigenvalues of $U$ on the unit circle. It is obvious that
$\Omega(U)=\Omega(U^{\dag})$ and $\Omega(U)=\Omega(T_1UT_2)$ for any
local operations $T_1,T_2$. In particular, $\Omega(U)=\Omega(U_d)$.

Up to a global phase, we have the following expression which is
given in Ref. \cite{CHN03}:
\begin{equation}\label{identity}
   U\sigma_y^{\otimes 2}U^T\sigma_y^{\otimes 2}=(u_A\otimes u_B)U_d^2(u_A^{\dag}\otimes u_B^{\dag})
\end{equation}
where the transpose "T" is taken with respect to the computational
basis $\{\ket{00},\ket{01},\ket{10},\ket{11}\}$ and
$\sigma_y^{\otimes 2}=\sigma_y\otimes \sigma_y$. Noticing that
$\Theta(A)=\Theta(X^\dagger AX)$ for any unitary $X$, we have
\begin{equation}\label{identy1}
\Omega(U)=\Theta(U\sigma_y^{\otimes 2}U^T\sigma_y^{\otimes 2}).
\end{equation}
The above equation will paly a key role in the proof of Lemma
\ref{lemma2} below as it provides a transparent connection between
$\Omega(\cdot)$ and $\Theta(\cdot)$.

It is easy to verify that $\Omega(U)=0$ if and only if $U$ is a
local operation or locally equivalent to Swap operation. Thus
$\Omega(U)>0$ if and only if $U$ is entangling \cite{BDDG+02}, i.e.,
$U$ can create entanglement from some unentangled input (without the
use of auxiliary systems).

Suppose now we are given a $2\otimes 2$ unitary $U$, and our purpose
is to simulate some perfect entangler using $U$ and with local
unitary operations as free resources. Clearly, if $U$ is not a
perfect entangler, then we need to apply $U$ more than one time. We
also require that $U$ is entangling, i.e., $\Omega(U)>0$. Otherwise
$U$ cannot create entanglement from any unentangled input. As any
entangling unitary $U$ together with local unitary operations is
universal \cite{BDDG+02}, there always exist a finite $N$ and a
sequence of local unitary operations $\{u_i\otimes v_k:k=0,\cdots,
N\}$ such that $(u_0\otimes v_0)U(u_1\otimes v_1)U\cdots (u_{N-1}\otimes
v_{N-1})U(u_k\otimes v_k)$ is a perfect entangler. A problem of great interest is to
determine the minimum of $N$.

Our main result is an analytical formula for the minimal number of
runs of $U$ required to simulate some perfect entangler. Most
interestingly, this formula is given in terms of $\Omega(U)$, say
$\lceil\frac{\pi}{\Omega(U)}\rceil$, and thus provides an
operational meaning of this quantity. More precisely, $\Omega(U)$
represents some kind of entanglement strength of $U$.

Before we present our main result Theorem \ref{theorem1}, let us
introduce some technical lemmas. They are also interesting in their
own right.

\begin{lemma}\label{lemma1}\upshape
A $2\otimes 2$ unitary operation $V$ is a perfect entangler if and
only if $\Omega(V)\geq \pi$. Furthermore,  if $\Omega(V)< \pi$, then
for every $\theta\in [\frac{\pi-\Omega(V)}{2},\frac{\pi}{2}]$, there
exists a state $|\psi\rangle$ such that $C(|\psi\rangle)=\sin\theta$
and $V|\psi\rangle$ is maximally entangled.
\end{lemma}

Remarks: The condition for perfect entangler has been obtained in
Ref. \cite{KC01} (implicitly) and latter in Ref.
\cite{ZVSW03}(explicitly). The condition for creating a maximally
entangled state from a pre-specified partially entangled state using
the given two-qubit unitary and local operations has been shown in
Ref. \cite{YSZG04}. A simplified proof together with a connection to
the distinguishability of unitary operations was then presented in
Ref. \cite{Che05}. Lemma \ref{lemma1} is of independent interest as
it gives an explicit condition $\Omega(V)\geq \pi$ which has not
been observed in the previous works. More importantly, Lemma
\ref{lemma1} characterizes all inial states $\ket{\psi}$ that can be
boosted into a maximally entangled by a single use of the given
two-qubit operation and local unitary operations.

\textbf{Proof:} We may assume that $\label{V} V=\sum_{k=1}^4
e^{i\lambda_k}\op{\Psi_k}{\Psi_k},$ where $\lambda_k$ are real
parameters. $V$ is a perfect entangler if and only if  that there is
a product state $|\phi\rangle$ such that $\ket{\Phi}=V|\phi\rangle$
is maximally entangled, or equivalently, $V^\dagger\ket{\Phi}$ is a
product state.

Without loss of generality, we may assume $\ket{\Phi}=\sum_{k=1}^{4}
\l_k|\Psi_k\rangle$ such that $l_k\in\mathcal{R}$. Then
$V^\dagger\ket{\Phi}=\sum_{k=1}^{4}
\l_ke^{-i\lambda_k}|\Psi_k\rangle$, and $
C(V^\dagger\ket{\Phi})=\sum_{k=1}^4 l_k^2 e^{-2i\lambda_k}=0$,  that
is, zero is contained in the convex hull of
$\{e^{-2i\lambda_k}:1\leq k\leq 4\}$. By a geometrical observation,
we know the above equation holds if and only if
$\Theta({V^{\dag}}^2)\geq\pi$. As the
$\Theta({V^{\dag}}^2)=\Theta(V^2)=\Omega(V)$ always holds, we have
$\Omega(V)\geq \pi$.

Suppose now $V$ is not a perfect entangler, i.e., $\Omega(V)<\pi$.
Again by a geometrical observation we know
$\cos\frac{\Omega(V)}{2}\leq |r| \leq 1$ for any point $r$ in the
convex hull of $\{e^{-2i\lambda_k}:1\leq k\leq 4\}$. Note that
$\theta\in [\frac{\pi-\Omega(V)}{2},\frac{\pi}{2}]$ implies
$\cos\frac{\Omega(V)}{2}\leq \sin\theta\leq 1$. By the intermediate
value theorem, there is a maximally entangled state $\ket{\Phi}$
such that $C(V^{\dag}\ket{\Phi})=\sin\theta$. Let
$|\psi\rangle=V^{\dag}\ket{\Phi}$, we have that
$C(|\psi\rangle)=\sin\theta$ and $V|\psi\rangle=\ket{\Phi}$ is
maximally entangled. \hfill $\blacksquare$\\

The following Lemma reveals a highly nontrivial property of
$\Omega(\cdot)$. A similar property for $\Theta(\cdot)$ has been
established in Ref. \cite{DFY07} and has been used as a key tool in
showing the optimality of the protocols for distinguishing unitary
operations.
\begin{lemma}\label{lemma2}\upshape
Let $U$ and $V$ be any two-qubit unitary operations such that
$\Omega(U)+\Omega(V)<\pi$. Then $\Omega(UV)\leq
\Omega(U)+\Omega(V)$.
\end{lemma}

\textbf{Proof:} We will employ the following properties of
$\Theta(\cdot)$ in the proof:

i) $\Theta(XUX^\dagger)=\Theta(U)$ for any unitary $X$. In
particular, $\Theta(UV)=\Theta(VU)$.

ii) If $\Theta(U)+\Theta(V)<\pi$, then $\Theta(UV)\leq
\Theta(U)+\Theta(V)$.

Item i) follows directly from the definition of $\Theta(.)$ and item
ii) was proven  in Ref. \cite{DFY07}.

Employing Eq. (\ref{identy1}), we have
\begin{eqnarray*}
\Omega(UV)&=&\Theta(UV\sigma_y^{\otimes 2}(UV)^T\sigma_y^{\otimes 2})\\
          &=&\Theta(UV\sigma_y^{\otimes 2}V^TU^T\sigma_y^{\otimes 2})\\
          &=&\Theta(V\sigma_y^{\otimes 2}V^TU^T\sigma_y^{\otimes 2}U)\\
          &=&\Theta(V\sigma_y^{\otimes 2}V^T\sigma_y^{\otimes 2}\sigma_y^{\otimes 2}U^T\sigma_y^{\otimes 2}U)\\
          &\leq &\Omega(V)+\Theta(\sigma_y^{\otimes 2}U^T\sigma_y^{\otimes 2}U)\\
          &=&\Omega(V)+\Theta(U\sigma_y^{\otimes 2}U^T\sigma_y^{\otimes 2})\\
          &=&\Omega(U)+\Omega(V),
\end{eqnarray*}
where the third and the fifth equality are due to item i), and the
first inequality is due to item ii) and Eq. (\ref{identy1}). \hfill
$\blacksquare$

To appreciate the power of Lemma \ref{lemma2}, let us consider the
following question: Given two entangling unitary operations $U$ and
$V$, how many uses of $U$ is necessary in order to simulate $V$
exactly, with local unitary operations as free resources.  For
simplicity, we assume that both $U$ and $V$ are not perfect
entanglers. Suppose now that $k$ runs of $U$ is sufficient to
simulate $V$, then there are local unitary operations $W_0,\cdots,
W_k$ such that
\begin{equation}V=W_0UW_1\cdots W_{k-1}UW_k.\end{equation}
Applying Lemma \ref{lemma2} to the above equation and noticing that
$k$ is an integer, we have
\begin{equation}
k\geq \lceil\frac{\Omega(V)}{\Omega(U)}\rceil.
\end{equation}
This is a lower bound of the necessary uses of $U$ to simulate $V$
with the assistance of local unitary operations.

Now we are ready to present our main result as follows.
\begin{theorem}\label{theorem1}\upshape
Let $U$ be a $2\otimes 2$ imprimitive unitary operation, and let
$N(U)=\lceil\frac{\pi}{\Omega(U)}\rceil$. Then there is a sequence
of local unitary operations $X_0,\cdots, X_{N(U)}$ such that
$X_0UX_1UX_2\cdots X_{N(U)-1}UX_{N(U)}$ is a perfect entangler.
Furthermore, for any $k<N(U)$ and any sequence of local unitary
operations $X_0,\cdots, X_k$, $X_0UX_1UX_2\cdots X_{k-1}UX_k$ cannot
be a perfect entangler. Thus $N(U)$ is the optimal number of runs of
$U$ to simulate some perfect entangler.
\end{theorem}

\textbf{Proof:} We first prove the second part. We will show that if
$k<N(U)$, then for any local unitary operations $X_0,...,X_k$,
$X_0UX_1\cdots X_{k-1}UX_k$ is not a perfect entangler. By Lemma 1, it
is sufficient to show that
\begin{equation}\label{Eq6}
   \Omega(X_0UX_1\cdots X_{k-1}UX_k)<\pi.
\end{equation}
Applying Lemma \ref{lemma2} $(k - 1)$ times, we have
\begin{equation}
   \Omega(X_0UX_1\cdots X_{k-1}UX_k)\leq \sum_{j=0}^{k-1}\Omega(X_jU)+\Omega(X_k)\leq k{\Omega(U)},
  \end{equation}
where we have used the fact that $\Omega(U)=\Omega(XU)$ for any
local unitary operation $X$. Now Eq. (\ref{Eq6}) follows from the
fact that $k\leq N(U)-1$ and $(N(U)-1)\Theta(U)<\pi$.

To prove the first part, we only need to show that by $N(U)$ times
of $U$ and local unitary operations we can transform a product state
into an maximally entangled state. For simplicity and without loss
of generality, we assume that $U_d=\sum_{k=1}^{4}
e^{i\lambda_k}\op{\Psi_k}{\Psi_k}$, and $\lambda_1\geq\lambda_2\geq
\lambda_3\geq \lambda_4\geq 0$. If $U$ is a perfect entangler, the
result trivially holds. If $U$ is not a perfect entangler, then
$\Omega(U_d)=2(\lambda_1-\lambda_4)<\pi$.  Let
$|\varphi\rangle=\frac{1}{\sqrt2}(\ket{\Psi_1}+i\ket{\Psi_4})$ be a
product input state and $N=N(U_d)$, then
$$C(U_d^{N-1}|\varphi\rangle)=\sin\frac{(N-1)\Omega(U)}{2}.$$
Note further that
$$\frac{(N-1)\Omega(U_d)}{2}\in
[\frac{\pi-\Omega(U_d)}{2},\frac{\pi}{2}].$$ It follows from Lemma
\ref{lemma1} that there is a state $\ket{\psi}$ such that
$C(\ket{\psi})=\sin\frac{(N-1)\Omega(U_d)}{2}$ and $U_d|\psi\rangle$
is maximally entangled. Applying the fact that for two-qubit states
$|\varphi_1 \rangle$ and $|\varphi_2\rangle$, $C(|\varphi_1
\rangle)=C(|\varphi_2 \rangle)$ if and only if there is a local
unitary operation $W$ such that$W|\varphi_1
\rangle=|\varphi_2\rangle$, we can choose local unitary operations
$W_1$ and $W_N$ such that $W_1|\alpha\beta\rangle=|\varphi\rangle$
and $W_NU_d^{N-1}|\varphi\rangle=|\psi\rangle$. Then
$U_dW_NU_d^{N-1}W_1|\alpha\beta\rangle$ is maximally entangled.
Thus, $U_dW_NU_d^{N-1}W_1$ is a perfect entangler, which implies
$N(U)$ runs of $U$ together with local unitary operations can
realize some perfect entangler. \hfill $\blacksquare$

The above theorem also presents the minimal number of runs of a
two-qubit entangling unitary operation to create a maximally
entangled state from some unentangled input. It is straightforward
to generalize this result to the case when the initial state is only
partial entangled. Let $|\tau\rangle$ be a two-qubit state and
$C(|\tau\rangle)=\sin\theta$, where $\theta\in [0,\pi/2)$.  There
exists a two-qubit unitary $V$ such that $\Omega(V)=2\theta$ and
$|\tau\rangle=V|00\rangle$. If $W_kUW_{k-1}U\cdots UW_1UW_0|\tau\rangle$
is maximally entangled, then $W_kUW_{k-1}U\cdots UW_1UW_0V$ is a perfect
entangler. It follows from Lemmas \ref{lemma1} and \ref{lemma2} that
\begin{equation}
k\geq\lceil\frac{\pi-2\theta}{\Omega(U)}\rceil.
\end{equation}
On the other hand, employing similar techniques as above, we know
$N(U,|\tau\rangle)=\lceil\frac{\pi-2\theta}{\Omega(U)}\rceil$ uses
of $U$ are sufficient to create a maximally entangled state. Now
it's easy to calculate the maximal reachable entanglement by a
single use of $U$ from $|\tau\rangle$, if $N(U,|\tau\rangle)>1$,the
final state is with concurrence less than or equal to
$\sin(\theta+\frac{\Omega(U)}{2})$; otherwise it is $1$, As shown in
Ref. \cite{YSZG04}.

Many interesting problems remain open. For instance, given a
$2\otimes 2$ unitary operations, what is the minimal number of runs
of $U$ in order to create maximal entanglement between Alice and
Bob. Here we assume Alice and Bob are far from each other and they
can perform arbitrary local operations and communicate classical
information with each other. Note that Alice and Bob now may prepare
entangled states locally. Thus any unitary locally equivalent to
Swap now can generate two maximally entanglement states by a single
run. It would also be interesting to generalize these results to
higher-dimensional case, where the situation would be very
different. Actually, from Ref. \cite{LSW05} we know that for some
$2\otimes 3$ unitary $U$, a single use of $U$ can create two copies
of $2\otimes 2$ maximally entangled states. However it requires at
least two uses of $U$ to disentangle two copies of $2\otimes 2$
maximally entangled states previously shared between Alice and Bob.

We are grateful to M. Y. Ye for pointing out useful references. This
work was partly supported by the Natural Science Foundation of China
(Grant Nos. 60736011, 60702080, and 60621062), the Hi-Tech Research
and Development Program of China (863 project) (Grant No.
2006AA01Z102), and the National Basic Research Program of China
(Grant No. 2007CB807901).

\end{document}